# A Comprehensive Review of Smart Cities Components, Applications, and Technologies Based on Internet of Things

**S. Aslam**
Dept. of Computer Science
*University of Gujrat, Gujrat, Pakistan*
Email: samiaaslam479@gmail.com

**H. Sami Ullah**
Dept. of Computer Science
*University of Gujrat, Gujrat, Pakistan*
Email: hamzasamiullah@gmail.com

## Abstract

Smart city technology is making cities more effective which is necessary for the rapid growth in urban population. With the rapid increase in advanced metering infrastructure and other digital technologies, Smart cities have become smarter with efficient electronic devices and embedded sensors based on the Internet of Things (IoT). This paper provides a comprehensive review of the smart cities concept with its components and applications. Moreover, technologies of IoT used in smart cities infrastructure and some practically implemented smart cities in the world are mentioned as exemplary implementations. Some open issues and future directions concluded the paper.

## 1. Introduction

The Internet of Things (IoT) is a contemporary research area in the modern world. IoT network is seen as a network of connected devices, sensors and actuators that are capable of connecting devices to the internet to send data and share information across the network [1]. With the rapid increase in the population of urban areas, cities with modern infrastructure and digital services are the necessities of residents. There will come many challenges with the increase in population in urban areas. To meet and reduce these challenges, a new concept and technology "Smart City" is emerging. The smart city provides and manages all daily life facilities i.e. Health, Retail system, water system, waste management system, transport, public safety measures, parking system, agriculture, energy and home solutions more efficiently and intelligently [2].

This smart city is connected with IoT infrastructure through sensor networks and remote devices for monitoring and proper working. Billions of IoT devices are operating under the whole smart city vision. The smart city architecture is correlated with IoT infrastructure for processing and delivering desired services to the end-users [3]. The paper is structured as follows: Section 1 introduces IoT and smart cities concepts with their basic architectures. Section 2 demonstrates smart cities emergence with various researches done in various components of smart city architecture. Section 3 explains various wireless technologies of IoT for smart cities. Section 4 describes some exemplary implemented smart city descriptions across the world. Some open research and implementation issues and future directions are given in section 5. The paper is concluded in section 6.

## 1. Background
### 1.1. Internet of Things (IoT)

Internet of Things (IoT) vision with all physical objects which are uniquely identified using RFID sensors has extended to the connected things anywhere and at any time. The term IoT was originally founded by MIT Auto-ID Center in 1999 [4]. Automatic Identification (Auto-ID) used for automatic data identification via bar code readers, Radio Frequency Identification (RFID), magnetic readers and optical memory cards. These technologies are used to automate any network, reduce errors and increase efficiency [5]. Therefore, the initial vision of IoT technology used RFID tags to uniquely identify and track physically connected objects. RFID tags were initially expensive but today the prices of individual RFID tags have decreased and made the adoption technically possible and increase economic feasibility [6]. Nowadays, the concept of IoT has evolved to provide insight into realizing a global infrastructure of the interconnectivity of physical and virtual objects. The new advancements in these technologies have extended the vision of IoT by encompassing other sensor networks.

The IoT architecture is based on three layers such as the perception layer, network layer, and application layer as shown in figure 1 with a brief description of the functionalities of each layer. The perception layer is the lowest layer of IoT architecture and it collects and perceives data and useful information from things connected in the network with the help of connected devices such as RFID [7], surveillance cameras and WSN. Data and information transfer from the perception layer to the application layer needs short-range networking and communication devices. The network layer provides these networking technologies such as WiFi, Bluetooth, Zigbee, 4G LTE, and z-wave to connect with wired or wireless connections to the application layer. The last application layer receives all the information from the perception layer via the network layer and processes that data into some useful and required output. With the layered architecture of IoT, Power distribution and management techniques can perform in a more efficient way.

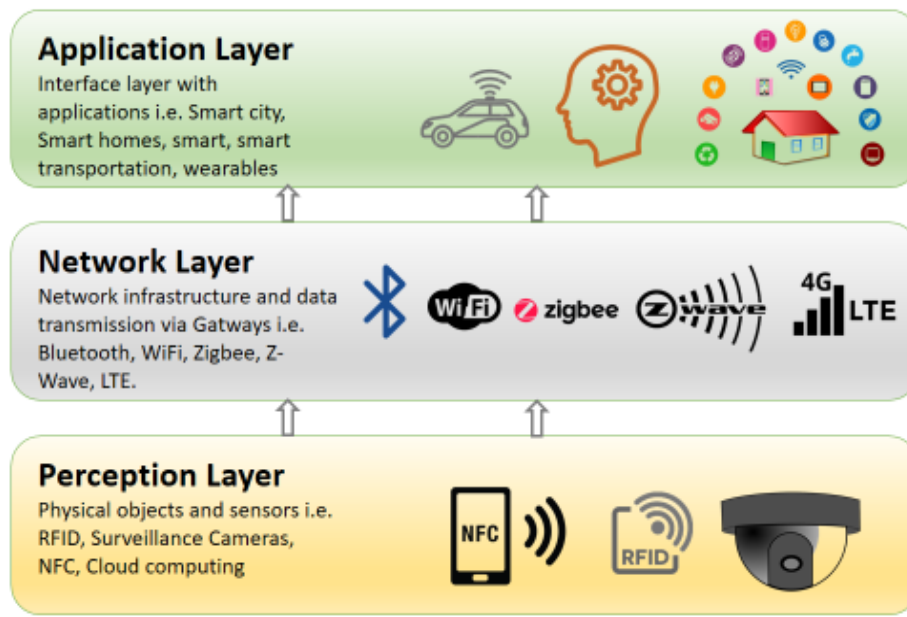

*Figure 1: Layers of IoT*

## 1.2. Smart Cities

Smart cities incorporate information and communication technologies (ICT) to enable automation and enhance the quality of urban living standards. Smart cities use integrated intelligent solutions to optimize the city infrastructure and provides responsive governance to engage the residents of the city in the management of their city environment. In [8] a smart city architecture based on IOT infrastructure is proposed with multi-levels approach. Raw data is collected in level 1 and data processing is done with the Resource Description Framework (RDF) format in level 2. Web Ontology Language, Dempster Shafer and SPARQL query language are used in level 3 and different web applications can be used for data control and alert in level 4. A smart building is proposed in [9] which is an essential implementation of the smart city. The main goal of SB is to provide the maximum comfort level of residences. For this purpose IOT base infrastructure is proposed which will collect, analyze and present real-time data.

A brief survey of architecture and technologies for an urban IoT for building a smart city is discussed in [10] where web service approaches, link layer technologies and devices are explained for an urban IoT architecture development. Then a practical implementation of discussed urban IoT architecture is discussed with Padova Smart City project. A CoT-based platform with an IoT network is proposed for a smart city and a VITAL project is presented as CoT-based architecture by [11]. Sensors and interconnected objects are used among multiple IoT platforms. VUAIs (Virtualized Universal Access Interfaces) are implemented for heterogeneous objects in VITAL. UWB (Ultrawideband) technology is proposed in [12] for IoT applications in a smart city.

## 2. IoT Potential Applications for Smart Cities

As there is a wide range of applications and services in the smart city paradigm and these services are responsible for providing better facilities to the citizens. Here an overview of smart city services and potential uses is described according to relevant researches done in the various components of smart city infrastructure.

## 2.1. Smart Homes

As the Internet of Things revolutionized the way we live and work, smart homes are filled with connected devices to make lives more convenient and easier. A lot of research has been conducted on home automation systems in recent years**.** In [13] a Home automation system is proposed for controlling all the home appliances and other electronic devices through the metering system and a website connected with it. Billing and dealership are also controllers through the website. AtHome is a system which controls the stationary wireless chargers. First, it calculates the amount of energy required for the IoT device then it will provide energy at the time of requirement automatically [14]. A simple smart home implementation is proposed in [15] to provide cost-effectiveness by making it dynamically useable. For example, if a new plug and play security camera are embedded with others then all other cameras will automatically share the security picture with it. The study [16] based on the survey about the people thinking for the IoT. The main parameter used in this study is enjoyment, compatibility and connectedness, control ease of use, usefulness, attitude and cost. Result's shows that people feel really good to use IOT products but little bit un-acceptance is shown with respect to the cost of these products. The design of wireless flood sensor for detecting water leakage and water on home floors is proposed in [17] where a wireless sensor network with sensor nodes is used to gather data and a control center and siren actuator node is used to process data and to warn the users about the water level. In [18] smart home adoption and diffusion are analyzed with different users with different behaviors of life. For analyzing adoption an extended technology acceptance model is used and for analyzing the spread of smart homes a multivariate probit model (MVP) was used. After analyzing this study suggests some necessary marketing strategies for a higher rate of the smart home market.

## 2.2. Smart Waste Management

The smart system for waste materials management is essential for the hygiene of the citizens. A garbage collection system is proposed in [19] where sensors are used in waste bin which is connected with GSM system and it will provide message notification to the garbage collectors when the waste bin reaches its maximum value. It will reduce time and cost and will increase greenery. In another paper a smart waste collection and management system is on real-time data using an IOT prototyping with sensors is proposed. The nearest-neighbor shortest path algorithm is used for checking nearest optimized routes and the shortest-path spanning tree algorithm is used for finding the distance of waste bins and work areas. Then waste in sensing algorithm is used for checking the level of waste bins and intelligent analysis is performed for the prediction of future waste bins load (Smart waste management using the Internet of Things (IoT). In [20] garbage and waste material management are done by using the idea of using minimal dustbins in a smart way with ultrasonic sensors with an alert beep and RFID technique is used for confirmation after emptying dustbin. To monitor remotely an android system is also attached with a web server for receiving notifications. In this paper accumulation and disposition of waste material problem is resolved with the help of ultrasonic sensors which tells the status of the waste containers to full, half or empty state and TCP/IP protocol is used to send sensed data over the web page and Google maps are used to identify the location of the containers [21].

## 2.3. Smart Traffic and Parking Management

While talking about services in a smart city, the importance of smart traffic and smart parking system cannot be neglected. In [22] impact of accidents on the traffic flow is monitored by using various sources of data and a system is proposed to store and process traffic data using the IoT and cloud technologies in the combine. The implementation is done only using Edmonton, Alberta Canada's Traffic data. In another study, a smart traffic light management system is proposed which uses a microcomputer with a transceiver connected to the server based on IoT. A wireless network is also used to connect the car computers to send cross intersection information on the road. Maximum Boolean derivative is used for controlling traffic on the roads [23]. In [24] an IoT based smart parking system with cloud integration is presented. Passive infrared and ultrasonic sensors are used as parking sensors and the Raspberry pi processing unit is used between cloud and sensors. MQTT message protocol is used for sending messages and a mobile application is developed in Apache Cordova and Angular js framework written in JavaScript. In another similar study, a smart parking system based on IoT is proposed for saving the time of people searching for parking slots. An embedded controller is used for user authentication while entering the parking lot. Parking meter with ultrasonic sensors is used for checking parking status and an LED detector is used for indicating the status free or reserved. An alarm IC module and camera are used for checking proper parking. A mobile application is used by users to communicate with the whole system. Apache Cordova and Angular js are used for developing applications with Javascript for Android and IOS users [25]. In [26] cloud-based car parking system is proposed which could be used in smart cities in the IoT paradigm. For this purpose, three-layer architecture is used 1. Sensor layer 2. Communication Layer 3. Application Layer. All these are connected through the cloud-based architecture to follow the anytime and anywhere communication concept.

## 2.4. Smart Water Management

Among all services and applications in smart city, smart water system and management of the distribution of water in urban and rural areas is very important. Several studies are conducted on the smart water management system for drinking and for agriculture as well. In [27] the importance and need of IoT based smart water distribution system is presented and discussed in detail. Quality parameters i.e. quality, quantity, technological and topological parameters for water are discussed and IoT application in the smart water system, smart gardening and smart irrigation is also described.

In another study [28] smart IoT based water distribution and monitoring systems are proposed for urban and rural areas of Pakistan. For monitoring, water quality sensors with a controller are used. Water distribution is controlled with IoT smart valves. In rural areas, the actuator system is proposed for monitoring and for supplying water in a proper flow. In this paper smart management of water for improving the quality of irrigation in agriculture. SWAMP architecture based on IoT infrastructure is presented in this paper. An IoT testbed is designed for FIWARE platform implementation [29]. For smart agriculture in this paper smart agricultural system based on IOT technology with solar supply is proposed. DHT11 is used for sensing the humidity and temperature of the fields. Moisture, water and rain sensors are used for checking the condition of crops and the PIR sensor is used for detecting movement of any person or thing in the fields. The Arduino platform is used for data collection. GSM sim900A is used for mobile communication to the farmers [30].

### 2.5. Smart Grids

Smart city perception also addresses the continuously arising issues of energy generation and consumption in urban cities. The energy infrastructure in a city is the most important feature as all other features in the city need the energy to function properly. Smart grids [31] modernized the traditional power system through self-maintaining designs, automation, remote monitoring and established microgrids. A smart grid infrastructure informs the consumers about their energy consumption and cost and provides a secure and reliable infrastructure with distributed energy sources. Smart cities are dependent on smart grids to ensure the resilient energy supply to other functions in the smart cities.

## 3. Technologies for IoT in Smart Cities

The network of IoT has a broadband connection of connected devices over the network with various communication protocols [32], [33]. Objects and devices in the IoT network are observed, measured and inferred to change the environment of smart city. IoT architecture is enabled by the development of different communication technologies and protocols. Some well-known technologies of IoT networks for smart cities applications are Radio Frequency Identification (RFID) [34], Wireless Sensor Networks (WSN) [35], IPV6 Addressing [36], and Middleware technologies [37].

## 4. Smart Cities in the World

With the rapid increase of information and communication technologies, Smart cities applications are changing the urban growth and their policy-making strategies. Here we describe some example projects of smart cities applications across the world with an enhanced experience of individuals, organizations and society in table 1.

*Table 1: Some projects of Smart City*

| Ref | Implementation and Outcomes |
|---|---|
| [38] | A connected public lighting system within smart city was implemented in Amsterdam with enhanced smart meters and smart controllers to calculate actual energy consumption with 80 percent of energy saving. |
| [38] | In New York to overcome violent crime, a City24/7 platform is created for the local citizens' awareness with citywide sensing smart screens and their communication capabilities. |
| [39] | In South Korea, the potential for implementing a good communication infrastructure of connected institutions, organizations, individuals, and government for sustainable urban development. |
| [38] | France investigated the potential to implement the Internet of Energy (IoE) and the economic model of IoE is tested and validated. Four services of smart cities were |

| | established such as smart lighting, smart circulation, smart environment monitoring, and smart waste management system. |
|---|---|
| [40] | Padova Smart City project was started in Padova with the collaboration of the University of Padova and the municipality of the city for running a city with smart automation. |
| [41] | Smart parking management system |

## 5. Open Issues and Future Trends

Smart cities have become an important mission for developed and developing countries with digital technologies to build an optimized and sustainable infrastructure for citizens. While making cities smarter need deployment of smart and complex infrastructure for implementing these digital technologies. With numerous connected devices, a smart city has to hire professionals to check and balance the whole infrastructure. However insufficient funds provided by the government are a big challenge and the government should create appropriate revenues for smart cities implementations and testing. Lack of skilled professionals is also another considerable challenge to maintain the smart city setup. The government should hire technical experts and professional designers for smart city project implementations. Consistent network availability is another major requirement in the smart environment of the city to gather data and information from sensors and to process them at a high speed of the internet. The reliable internet connectivity in smart city architecture is mandatorily taken to account for efficient data processing in a real-time environment. Cybersecurity Challenges are also important to consider by the government authorities and IT specialists to make real-time data secure and resilient to different attacks [42]. While we agree upon the smart technologies in smart cities to make our lives easier than ever especially in urban areas, but technology always needs to implement and maintain the structure appropriately and in a highly secure manner. The effects of technology must be considered with the provided solutions in smart cities to ensure reliability and accuracy.

## 6. Conclusion

Internet of Things (IoT) has become a more advanced technology with the rapid advancements in the vision to interact and communicate with physical objects at any time and anywhere in smart cities infrastructure. In this paper, we reviewed IoT architecture integrated with smart cities and various smart cities applications according to different proposed and implemented researches that are briefly explained. Wireless technologies used in IoT network while working with smart cities are also explained. Some open research issues and future trends are suggested for future researchers.

[24] A. Khanna and R. Anand, "6-IoT based smart parking system," in *2016 International Conference on Internet of Things and Applications (IOTA)*, Pune, India, 2016, pp. 266–270, doi: 10.1109/IOTA.2016.7562735.

[25] D. Issrani and S. Bhattacharjee, "7-Smart Parking System Based on Internet of Things: A Review," in *2018 Fourth International Conference on Computing Communication Control and Automation (ICCUBEA)*, Pune, India, 2018, pp. 1–5, doi: 10.1109/ICCUBEA.2018.8697348.

[26] Z. Ji, I. Ganchev, M. O'Droma, L. Zhao, and X. Zhang, "8-A Cloud-Based Car Parking Middleware for IoT-Based Smart Cities: Design and Implementation," *Sensors*, vol. 14, no. 12, pp. 22372–22393, Dec. 2014, doi: 10.3390/s141222372.

[27] V. Radhakrishnan and W. Wu, "9-IoT Technology for Smart Water System," in *2018 IEEE 20th International Conference on High Performance Computing and Communications; IEEE 16th International Conference on Smart City; IEEE 4th International Conference on Data Science and Systems (HPCC/SmartCity/DSS)*, Exeter, United Kingdom, 2018, pp. 1491–1496, doi: 10.1109/HPCC/SmartCity/DSS.2018.00246.

[28] S. Safdar, M. Mohsin, L. A. Khan, and W. Iqbal, "10-Leveraging the Internet of Things for Smart Waters: Motivation, Enabling Technologies and Deployment Strategies for Pakistan," in *2018 IEEE SmartWorld, Ubiquitous Intelligence Computing, Advanced Trusted Computing, Scalable Computing Communications, Cloud Big Data Computing, Internet of People and Smart City Innovation (SmartWorld/SCALCOM/UIC/ATC/CBDCom/IOP/SCI)*, 2018, pp. 2117–2124, doi: 10.1109/SmartWorld.2018.00354.

[29] C. Kamienski *et al.*, "11-Smart Water Management Platform: IoT-Based Precision Irrigation for Agriculture," *Sensors*, vol. 19, no. 2, p. 276, Jan. 2019, doi: 10.3390/s19020276.

[30] "12-Smart Agriculture Monitoring and Protection System Using IOT," vol. 2, no. 12, p. 3, 2019.

[31] M. Masera, E. F. Bompard, F. Profumo, and N. Hadjsaid, "Smart (Electricity) Grids for Smart Cities: Assessing Roles and Societal Impacts," *Proc. IEEE*, vol. 106, no. 4, pp. 613–625, Apr. 2018, doi: 10.1109/JPROC.2018.2812212.

[32] L. Atzori, A. Iera, and G. Morabito, "The Internet of Things: A survey," *Comput. Netw.*, vol. 54, no. 15, pp. 2787–2805, Oct. 2010, doi: 10.1016/j.comnet.2010.05.010.

[33] T. Robertson and I. Wagner, "CSCW and the Internet of Things," in *ECSCW 2015: Proceedings of the 14th European Conference on Computer Supported Cooperative Work, 19-23 September 2015, Oslo, Norway*, Cham, 2015, pp. 285–294, doi: 10.1007/978-3-319-20499-4_15.

[34] N. Deng, "RFID Technology and Network Construction in the Internet of Things," in *2012 International Conference on Computer Science and Service System*, 2012, pp. 979–982, doi: 10.1109/CSSS.2012.248.

[35] "Convergence of MANET and WSN in IoT Urban Scenarios - IEEE Journals & Magazine." [Online]. Available: https://ieeexplore.ieee.org/abstract/document/6552998. [Accessed: 04-Feb-2020].

[36] T. Savolainen, J. Soininen, and B. Silverajan, "IPv6 Addressing Strategies for IoT," *IEEE Sens. J.*, vol. 13, no. 10, pp. 3511–3519, Oct. 2013, doi: 10.1109/JSEN.2013.2259691.

[37] "Middleware for Internet of Things: A Survey - IEEE Journals & Magazine." [Online]. Available: https://ieeexplore.ieee.org/abstract/document/7322178. [Accessed: 04-Feb-2020].

[38] S. Mitchell, N. Villa, M. Stewart-Weeks, and A. Lange, "Connecting People, Process, Data, and Things To Improve the 'Livability' of Cities and Communities," p. 21.

[39] E. Strickland, "Cisco bets on South Korean smart city," *IEEE Spectr.*, vol. 48, no. 8, pp. 11–12, Aug. 2011, doi: 10.1109/MSPEC.2011.5960147.

[40] A. Cenedese, A. Zanella, L. Vangelista, and M. Zorzi, "Padova Smart City: An urban Internet of Things experimentation," in *Proceeding of IEEE International Symposium on a World of Wireless, Mobile and Multimedia Networks 2014*, 2014, pp. 1–6, doi: 10.1109/WoWMoM.2014.6918931.

[41] G. P. Hancke, B. D. C. e Silva, and J. Hancke, "The Role of Advanced Sensing in Smart Cities," *Sensors*, vol. 13, no. 1, pp. 393–425, Jan. 2013, doi: 10.3390/s130100393.
8